# PHOTODISINTEGRATION OF $^4$He NUCLEUS INTO p$^3$H AND n$^3$He CHANNELS IN THE POTENTIAL CLUSTER MODELS

S.B. Dubovichenko

Processes of direct cluster photodisintegration of $^4$He nucleus into p $^3$H and n $^3$He channels are considered on the base of two-cluster potential models. Intercluster interactions being in agreement with elastic scattering phase shifts and characteristics of the bound state of a nucleus have, in some cases, forbidden states. The scattering phase shifts and potentials are separated in the channels with minimum spin on the base of Young schemes and isospin states.

Earlier it was shown [1,2] that there is orbital symmetry mixing and sometimes isospin mixing in the continuum of light N $^2$H, p $^3$H, n $^3$He, $^2$H $^3$He and $^2$H $^2$H clusters in the minimum spin channels while the ground states (GS) of $^3$H, $^3$He and $^4$He are usually considered as pure. Interactions obtained on the base of scattering phase shifts in the minimum spin channels depend effectively on two Young schemes and can not be used for an analysis of bound state (BS) characteristics in the two-cluster model. It is necessary to extract pure component which can be applied to GS analysis from such interactions. In this case results of calculation of the GS characteristic depend mainly on the probability of nucleus clusterization in the channels under consideration.

Regarding the lightest cluster systems experimental mixed phase shifts may be represented as a half-sum of pure phase shifts with some specific Young schemes [1,2]. It is usually assumed that a pure phase shifts of another spin state or isospin pure system can be used for one of the pure phase shifts of a minimum spin channel. In this case, using the experimental phase shifts one can find easily a pure phase shifts of maximum symmetry and obtain a pure interaction.
Such N $^2$H, N $^3$H, N $^3$He and $^2$H $^2$H interactions were obtained in [1,2] and it was shown [2] that, in general, it is possible to describe correctly $^3$H, $^3$He and $^4$He bound state energy in cluster channels, an asymptotic constant, a charge radius and elastic Coulomb formfactor at low momenta transferred. The potential cluster model was used in the calculations [2] assuming that a nucleus consists of two structureless fragments with properties of some certain particles in a free state. A full antisymmetrization of the system wave function is not carried out but deep intercluster interactions in some spin channels contain forbidden states (FS). Due to this fact wave function of cluster relative motion at short range is not equal to zero as it occurs in case of core potentials but oscillates in the nucleus internal space. The phase shifts are in agreement with a generalized Levinson theorem and tend to zero at high energies.

Orbital symmetry mixing results in substantial differences between the lightest cluster systems and the $^2$H $^4$He, $^3$H $^4$He, $^3$He $^4$He and $^3$He $^3$H systems where the orbital states are pure and it becomes possible using phase shifts to obtain interactions with forbidden states enabling to describe some characteristics of $^6$Li, $^7$Li and $^7$Be [3] nuclei.

Note that the orbital Young scheme mixing in the minimum spin states is characteristic not only for the above-mentioned lightest cluster systems but for some heavier systems like N $^6$Li, N $^7$Li and $^2$H $^6$Li as well [4].

The calculation of differential cross sections of photoprocesses in N $^2$H, p $^3$H and $^2$H $^2$He channels were carried out for the lightest nuclei on the base of cluster models with FS potentials and orbital Young scheme separation [1]. The total cross sections for potential cluster models with an orbital Young scheme separation were considered in [5].

At first let us consider the state symmetry in pure systems where isospin T is equal only to 1 with $\{31\}_T$ isospin scheme. Spin states with S=1,0 are characterized by $\{31\}_S$ and $\{22\}_S$ Young schemes, correspondingly. Spin-isospin symmetry of a wave function which is a direct internal product of spin and isospin schemes [1,6] gives $\{4\}+\{31\}+\{22\}+\{211\}$ in ST=1,1 state and $\{31\}+\{211\}$ in ST=0,1 state. Total wave function symmetry is a direct internal product of spin-isospin and orbital symmetries. Possible orbital symmetries are determined on the base of the Littlewood theorem and are a direct external product of orbital schemes of sub-systems [1]. In case of 1+3 particles we have $\{1\} \times \{3\}=\{4\}+\{31\}$. The allowed orbital states are the states with conjugate symmetry to spin-isospin schemes. So, it becomes clear that at ST=1,1 $\{31\}$ orbital symmetry conjugated to $\{211\}$ is allowed. This $\{31\}$ orbital symmetry is allowed at ST=0,1, too. Thus, there are forbidden states with $\{4\}$ in triplet and singlet states and P-state has an allowed bound level with $\{31\}$.

Let now consider possible symmetries in p$^3$H and n$^3$He systems where spin and isospin can be equal to 0 and 1. It is clear that at T=1 and S=0,1 wave function structure coincide with the above-mentioned ones. So consider in detail the case of T=0 at S=0,1. Spin and isospin wave functions are characterized by $\{22\}$, $\{22\}$ and $\{31\}$ Young schemes, correspondingly. Spin-isospin symmetry at ST=1,0 is determined to be $\{31\}+\{211\}$ and coincide with ST=01 symmetry in p$^3$He and n$^3$H systems. Spin-isospin symmetry $\{1^4\}+\{22\}+\{4\}$ was obtained in case of ST=0,0. Possible orbital Young schemes were determined above. Hence, orbital symmetry $\{4\}$ is allowed and $\{31\}$ is forbidden. So isospin state at T=0 is characterized by two orbital Young schemes. And two schemes $\{4\}$ and $\{31\}$ corresponding to isospin states 0 and 1 are allowed at S=0.

It is shown in [1, 2] that scattering phase shifts $\delta_{l}LS$ can be represented for p$^3$H system as follows:

$$2\delta_{L0} = \delta^{\{4\}}_{L00} + \delta^{\{31\}}_{L01}, \quad 2\delta_{L1} = \delta^{\{31\}}_{L10} + \delta^{\{31\}}_{L11}$$

where $\delta_{LST}$ is isospin pure phase shifts. Hence to obtain pure singlet and triplet phase shifts at T=0 with $\{4\}$ and $\{31\}$ it is possible to use phase shifts with T=1 of singlet and triplet states of p$^3$He system with $\{31\}$. The pure phase shifts enable to carry out a parameterization of the isospin pure interaction potentials in p$^3$H and n$^3$He cluster channels.

To calculate the radiative capture cross sections in the long-wave approximation a well-known formula [7] was used:

$$\sigma_c(NJ) = \frac{8\pi}{\hbar^2 q^3} \frac{K^{2J+1}}{(2s_1+1)(2s_2+1)} \frac{\mu}{J[(2J+1)!!]^2} \frac{J+1}{J} \sum_{\substack{m_i, m_f, \\ m}} |M_{Jm}(N)|^2, \tag{1}$$

where N=E is an electric or N= M is a magnetic transition and $H_{Jm}(N)$ is electromagnetic operators in cluster model [5], J is a multiplicity, q is a wave number of cluster relative motion, $\mu$ is a reduced mass of a nucleus in the cluster model, K is a photon wave number. The photodisintegration cross sections can be determined using the detailed balancing principle.

Experimental phase shifts and cross sections of p$^3$He interaction are well-known for rather wide energy range and different measurements results are in a good agreement with each other [8, 9]. The substantial difference in the experimental results of phase shifts is only in

$P_1$ waves at S=0,1. Cross sections and phase shifts of p$^3$H system at 3-4 MeV are described in [10] and at higher energies - in [11]. There are two variants of phase shifts analysis at energies 4-15 MeV [11], the results of the both variants were used to obtain pure phase shifts [2]. The central Gaussian potentials of intercluster interaction used in [2] are the following:

$$V(r) = - V_0 \exp(-\alpha r^2) + V_c(r)$$

where $V_c$ is a point Coulomb interaction. In some cases a peripheral repulsive interaction + $V_1 \exp(-\beta r)$ was added to describe both positive S and negative D (at low energies) scattering phase shifts.

To calculate photo cross section E1 processes a transition between pure GS with T=0 and a singlet scattered wave were considered. If one supposes that the processes with isospin change $\Delta T=1$ [12] made the main contribution to the cross section then P potential of T=1 isospin pure singlet state of p$^3$He system should be used. If $\Delta T=0$ transitions are considered then a pure interaction from p$^3$H system should be used as P potential.

Interaction with $V_0 = 63.1$ MeV, $\alpha = 0.17$ Fm$^{-2}$ was obtained in [2] for the case of GS with T=0. The results of calculation of $^4$He GS energies, charge radius and asymptotic constant $C_w$ are listed in the Table together with the experimental data [13]. The asymptotic constant is determined by a standard method using Witteker functions. Results obtained for 72.5 MeV and 0.25 Fm$^{-2}$ potential [1] are included in the Tabl.1 for comparison. The charge radiuses were determined from the elastic Coulomb formfactors in the limits of a low momentum transferred [2].

Tabl.1. GS energies, $^4$He charge radius and asymptotic constants for pure potentials in p$^3$H and n$^3$He systems

| No. | p$^3$H | | | n$^3$He | | |
|---|---|---|---|---|---|---|
| | E, MeV | $C_w$, Fm | R, Fm | E, MeV | $C_w$, Fm | R, Fm |
| Calculated (2) | -19.82 | 4.6 (1) | 1.69 | -20.86 | 4.4 (1) | 1.70 |
| Calculated [1] | -19.79 | 3.7 (1) | 1.64 | -20.92 | 3.5 (1) | 1.67 |
| Experim. [13] | -19.815 | 4.2 (2) -5.2 (1) | 1.673(1) | -20.578 | 5.1 (4) | 1.673(1) |

A potential with $V_0$=15.0 MeV and $\alpha$ =0.1 Fm$^{-2}$ enabling to make a compromise between two different phase shifts analysis [8] (triangles) and [9] (dots, squares) was used as singlet P potential of p$^3$He scattering (Fig. 1a, solid line). To describe phase shifts [8] the potential depth should be reduced to 11 MeV and phase shifts [9] can be represented by an interaction with 17 MeV depth. These results shown in Fig.1a by dot and dash lines, correspondingly. Isospin pure S and D phase shifts with T=1 obtained using peripheral repulsive potential [2] with Vo=110 MeV, $\alpha$ =0.37 Fm$^{-2}$ and $V_1$=45 MeV, $\beta$ =0.67 Fm$^{-1}$ are represented in Fig.1a. by solid lines.

Pure phase shifts of p$^3$H system obtained from experimental phase shifts are shown in Fig.1b by dots. The calculated phase shifts with above potentials are shown by solid lines. Negative P phase shift may be parametrized by a repulsive potential with $V_0$=-8 MeV and $\alpha$ =0.03 Fm$^{-2}$. Dot line in even waves is used for the calculated phase shifts for the potential from [1].

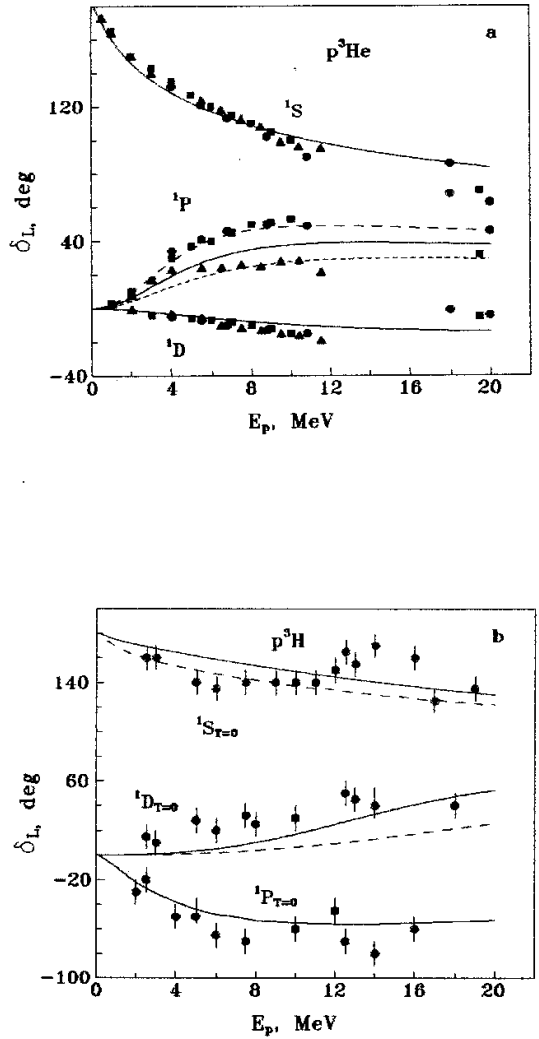

Fig.1. (a) - phase shifts of p$^3$He elastic scattering. (b) - points are for the pure phase shifts of elastic p$^3$H scattering with {4}.

Results of our calculations of the total cross section of $^4$He photodisintegration into p $^3$H channel at Δ T=1 transitions are shown in Fig. 2a by a solid line for the our GS potential and for the P-interaction with 15 MeV. The experiment data are from [14] (circles), [15] (triangles) and [16] (dots) and shown in Fig.2a. It is clear that the differences in the experimental data reaches 20-30 %. Results of the calculation for the potential [1] are presented by a dot line and are much lower than any experimental data. The results of assumption that there are Δ T=0 transitions are shown by a dash line which does not even reproduce the form of the experimental cross section.

The calculation results for $^4$He ( γ , n) $^3$He reaction with Δ T=1 presented in Fig.2b are compared with the experimental data from [16] (dots), [17] (triangles) and [18] (circles). The calculated curves are presented as above. If the potential depth in P wave is reduced to 11 MeV at the same geometry it is possible to describe data from [16, 17]. This cross section is shown in Fig.2b by a dash line. Phase shifts of such P potential are presented in Fig. 1a by a dot line and agree with the phase shift analysis data [8].

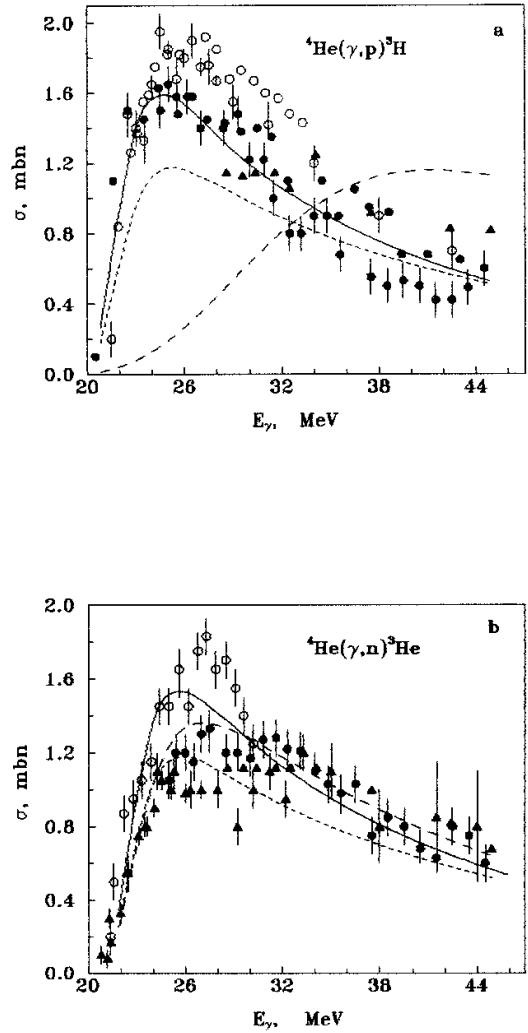

Fig.2. - total cross sections for photodisintegration of $^4$He nucleus.

Cross sections of E2 $^4$He($\gamma$, p) $^3$H process of transition from the ground state into singlet D wave are shown in Fig. 3a for transitions with $\Delta$ T=0 ( solid line) and $\Delta$ T=1 (dash line). A pure GS potential with T=0 of p$^3$H system (see Fig.1) was used for D wave in the first case and D wave peripheral repulsive interaction with T=1 of p$^3$He system was used in the second case. The experiment data are described in [14]. It is clear that the more acceptable results can be obtained for E2 processes if it is assumed that the $\Delta$ T=0 transitions ensure the main contribution. But in this case the calculated cross sections are lower than the experimental results.

The astrophysical S factor for E1 capture at low energies was considered for the above GS and P interaction 15 MeV depth potential. This S factor is shown in Fig. 3b. It is clear that despite the experimental errors it is possible to reproduce the experimental S factor [14-16] at low energy 0.7-3 MeV. A linear extrapolation of the S-factor to zero energy gives a value of about $10^{-3}$ keV b.

Thus, the potential cluster model based on the GS potential and a compromise P interaction enable to reproduce the photodisintegration cross section shape at E1 transitions with $\Delta$ T=1 for the both reactions considered. The calculated cross section values are in the range of ambiguity for different experimental results. In case P interaction depth is varied between 10-20 % and the results of calculation will better agree with any experimental data. Phase shifts of the P potentials changed in such a way are in the range of the experimental ambiguity of different phase shifts analysis.

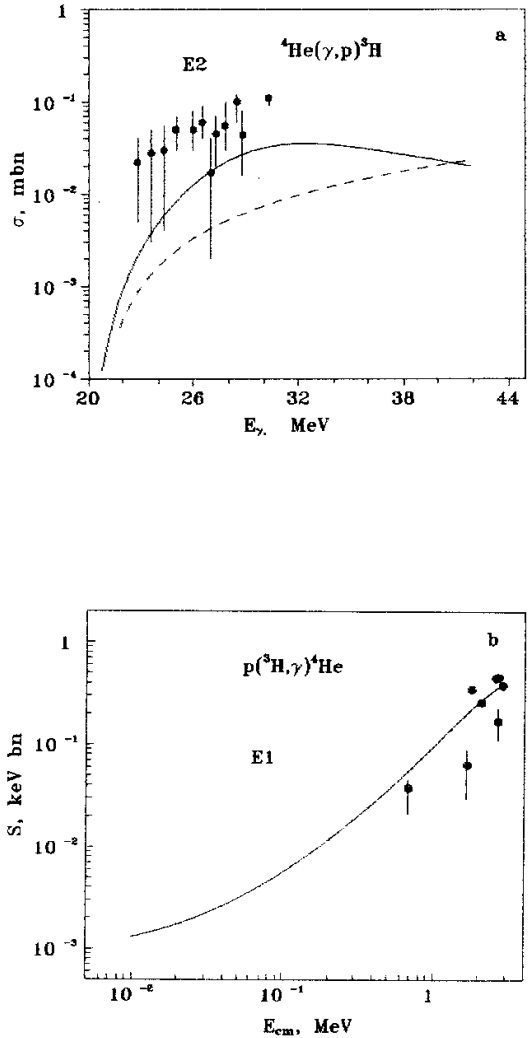

Fig.3. (a) - total cross sections for E2 photodisintegration of $^4$He nucleus. (b) - astrophysical S-factor for p$^3$H radiative capture at low energies.